\documentclass[aps, prd, showpacs, superscriptaddress, nofootinbib, twocolumn]{revtex4}
\usepackage{graphicx}\usepackage{amsmath}\usepackage{natbib}\usepackage{epstopdf}
\begin{document}
\title{Neutrino Constraints on the Dark Matter Total Annihilation Cross Section}

\author{Hasan Y{\"u}ksel}
\affiliation{Department of Physics, Ohio State University, Columbus, Ohio 43210}
\affiliation{Center for Cosmology and Astro-Particle Physics, Ohio State University, Columbus, Ohio 43210}

\author{Shunsaku Horiuchi}
\affiliation{Department of Physics, School of Science, University of Tokyo, Tokyo 113-0033, Japan}

\author{John F. Beacom}
\affiliation{Department of Physics, Ohio State University, Columbus, Ohio 43210}
\affiliation{Department of Astronomy, Ohio State University, Columbus, Ohio 43210}
\affiliation{Center for Cosmology and Astro-Particle Physics, Ohio State University, Columbus, Ohio 43210}

\author{Shin'ichiro Ando}
\affiliation{California Institute of Technology, Mail Code 130-33, Pasadena, CA 91125}

\date{17 December 2007}


\begin{abstract}
In the indirect detection of dark matter through its annihilation products, the signals depend on the square of the dark matter density, making precise knowledge of the distribution of dark matter in the Universe critical for robust predictions. Many studies have focused on regions where the dark matter density is greatest, e.g., the Galactic Center, as well as on the cosmic signal arising from all halos in the Universe.  We focus on the signal arising from the whole Milky Way halo; this is less sensitive to uncertainties in the dark matter distribution, and especially for flatter profiles, this halo signal is larger than the cosmic signal. We illustrate this by considering a dark matter model in which the principal annihilation products are neutrinos. Since neutrinos are the least detectable Standard Model particles, a limit on their flux conservatively bounds the dark matter total self-annihilation cross section from above. By using the Milky Way halo signal, we show that previous constraints using the cosmic signal can be improved on by 1--2 orders of magnitude; dedicated experimental analyses  should be able to improve both by an additional 1--2 orders of magnitude.
\end{abstract}


\maketitle


\section{Introduction}
Dark matter, an invisible substance originally proposed by Zwicky in the 1930s to explain the mass-to-light ratio of the Coma galaxy cluster~\cite{Zwicky:1933}, still evades revealing its true identity after many decades. Subsequent works have confirmed its pervasive existence, and its contribution to the critical energy density has been measured to be $\Omega_{\chi} \simeq 0.3$. The distribution of dark matter (DM) in the universe has also been thoroughly studied on a wide range of scales. Many DM candidates have been proposed, from axions to sterile neutrinos to weakly interacting massive particles (WIMPs); see recent reviews~\cite{Jungman:1995df,Bergstrom:2000pn,Bertone:2004pz}.  WIMPs are especially popular since they are among the best theoretically motivated non-baryonic DM candidates and also capable of producing the correct relic density. They can be detected indirectly through the astrophysical signatures of their annihilation products, detected directly in low background experiments underground, and produced at particle accelerators.

For indirect detection, the DM distribution is essential for predicting the detectability of the signals, as the annihilation rate depends on the annihilation cross section and the square of the DM number density (the DM mass density divided by the DM mass).  The Galactic Center (GC)~\cite{Bengtsson:1990xf,Berezinsky:1994wv,Bergstrom:1997fh,Bergstrom:1997fj,Bergstrom:2000bk,Cesarini:2003nr,Hooper:2004vp,Fornengo:2004kj,Horns:2004bk,Gondolo:1999ef,Ullio:2001fb, Gnedin, Ahn:2007ty,Bertone:2004ag}, intermediate mass black holes~\cite{Bertone:2005xz,Bertone:2006nq,Horiuchi:2006de,Brun:2007tn,Fornasa:2007ap}, dwarf satellite galaxies~\cite{Profumo:2006hs,Bergstrom:2005qk,Strigari:2006rd}, and nearby galaxies~\cite{Fornengo:2004kj,Evans:2003sc}, i.e., places where the dark matter is strongly concentrated, have been considered in the search for DM signals. Additionally, the diffuse cosmic signal due to all distant concentrations of DM has been utilized as well~\cite{Bergstrom:2001jj,Ullio:2002pj,Taylor:2002zd,Elsaesser:2004ck,Elsaesser:2004ap,Ando:2005hr,Oda:2005nv,Ando:2005xg,Horiuchi:2006de,Ando:2006cr}. 

In the Milky Way (MW), an attractive site for consideration is the GC, due to an expected increase in the DM density towards the centers of gravitational potentials~\cite{Moore:1999gc,Navarro:1995iw,Kravtsov:1997dp}, and possibly a spike around the central black hole~\cite{Gondolo:1999ef,Ullio:2001fb, Gnedin}.  However, the inner gradient of the halo profile is the most uncertain part. Although simulations generally suggest a steep cusp towards the halo center, some observations, such as the high microlensing optical depth, support much flatter profiles~\cite{Binney:2001wu}. In fact, there is no direct observational evidence that any nearby galaxy has a cusped DM profile. The small spatial resolution needed to settle the debate has not yet been reached in simulations. Another problem is astrophysical backgrounds. For example, astrophysical gamma-ray sources in the GC may limit the DM parameter space that can be probed with gamma rays~\cite{Zaharijas:2006qb}. So it may be preferable to study other regions of the Milky Way, where both the signal and backgrounds are more certain.

In this study, DM annihilation in the whole MW halo, as opposed to just the GC, is our primary focus. While the GC signal strongly depends on the uncertain cuspiness of the underlying halo profile, the signal averaged over larger regions of the halo is much less dependent on the details as long as the density normalization is chosen consistently with observed data, e.g., rotation curves~\cite{Klypin:2001xu}. This has been emphasized  by e.g., Refs.~\cite{Calcaneo-Roldan:2000yt,Stoehr:2003hf,Hooper:2007be}.

Using the annihilation signal arising from the halo means accepting a lower signal intensity, but the intensity of the astrophysical backgrounds is also lower. The halo signal is less subject to theoretical uncertainties compared to the GC or the cosmic signal, and can provide more robust results. Moreover, regardless of the parametrization, the halo always provides a minimal isotropic contribution that cannot be distinguished from any truly diffuse cosmic signal.  The halo isotropic component is increasingly dominant over the cosmic signal for flatter halo profiles.

In indirect detection, regardless of the identity of the DM particle, one would expect the most stringent constraint on the annihilation cross section from the most detectable annihilation products, and such gamma rays have long been sought. However, the branching ratios to produce gamma rays are small and uncertain, making it impossible to set a certain limit on the total annihilation cross section. Since neutrinos are much harder to detect, one may naively expect that they cannot yield any meaningful constraints. However, this is not the case. 

While both the gamma ray and neutrino flux predictions are very much dependent on WIMP model parameters, one can avoid these complications by adopting a more conservative perspective, following Beacom, Bell and Mack (BBM)~\cite{Beacom:2006tt}. If we assume that only Standard Model (SM) final states are produced, e.g., purely sterile neutrinos are not considered, then neutrinos and antineutrinos are the least detectable final states. Anything else will inevitably produce the much more visible gamma rays, leading to a stronger cross section limit. Thus, one can obtain a stringent upper limit on the total annihilation cross section by assuming that only neutrinos are produced as final states, and comparing the expected signal to the well-measured and modeled flux of atmospheric neutrinos, which is the dominant background.

We focus on the importance of the annihilation signal arising from the halo in the context of obtaining model-independent bounds on the DM annihilation cross section, as described above.  For very large annihilation cross sections, as in Ref.~\cite{Kaplinghat:2000vt}, halos would be significantly modified.  BBM have shown that such large cross sections are not possible~\cite{Beacom:2006tt}, since then the cosmic signal, which depends on the time-integrated annihilation rate of all halos, cannot be accommodated in comparison to the atmospheric neutrino spectrum.  In this study, we instead use the MW halo signal, which is generally larger and more certain, as we show below, to set an even stronger limit on total annihilation cross section. 

We first summarize the halo profiles and the model for annihilating DM in the halo and in cosmic sources.  We also identify the various components of the annihilation signal arising from the halo. To provide the most general results, we consider various profile shapes for the halo together with a dark matter model in which neutrinos are the only annihilation products.  Next, we compare the neutrino spectrum arising from annihilations to the atmospheric neutrino spectrum, and arrive at our constraints on the total annihilation cross section.  Our study shows that even the full-sky halo signal would lead to over an order of magnitude improvement on the cosmic signal limit on the cross section.  If the broad enhancement towards the GC direction is taken into account, further improvement is possible while remaining secure with respect to the halo profile uncertainties. 


\begin{figure}[t]
\includegraphics[width=3.25in,clip=true]{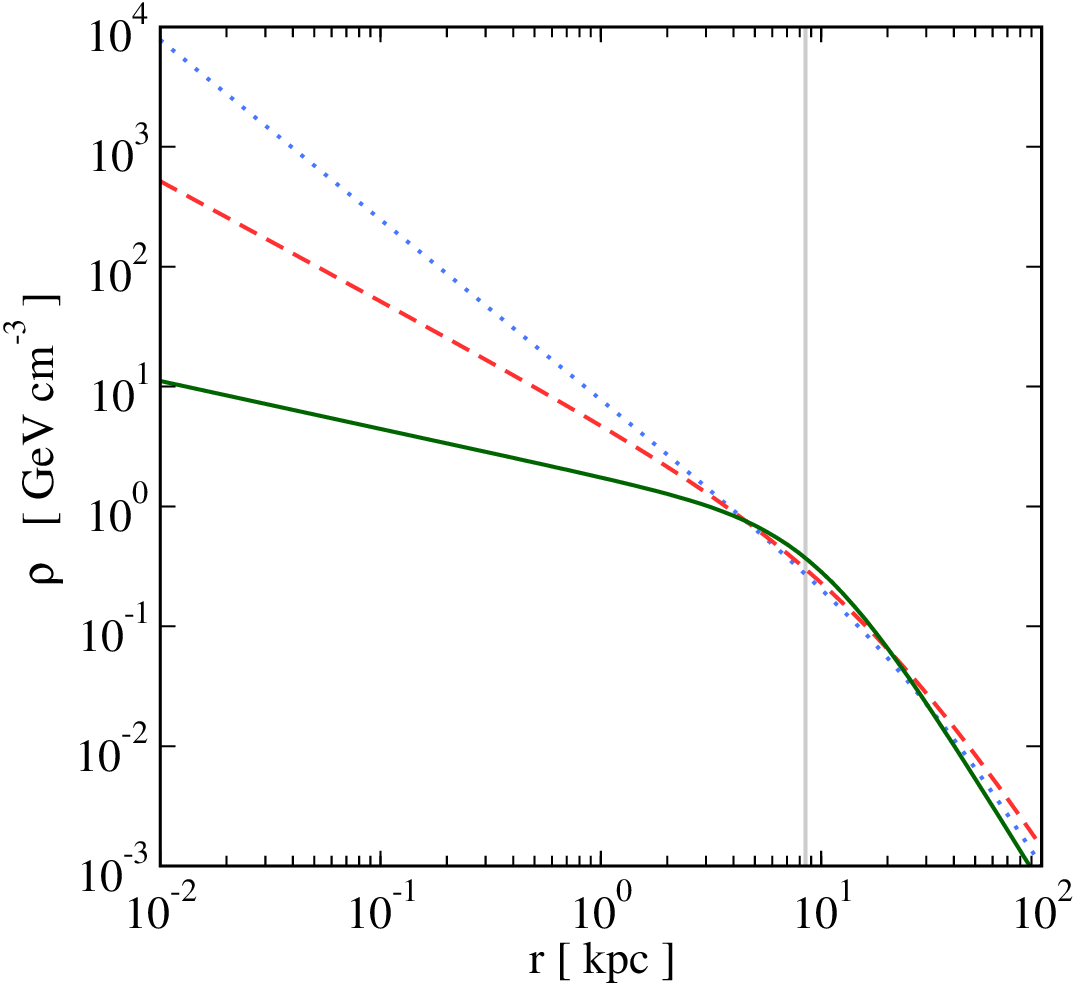}
\caption{The dark matter density versus radius for the three profiles considered (Moore, NFW and Kravtsov in order of dotted, dashed and solid lines), based on Eq.~(\ref{eq:fitting}). The normalizations are chosen to match the large scale properties of the Milky Way. The vertical gray line indicates the solar circle distance, $R_{sc}\simeq 8.5$~kpc. See Table~I for the other parameters.
\label{fig-1}}
\end{figure}

\section{Annihilating Dark Matter Model}
Cold dark matter simulations show that DM assembles hierarchically in halos and successfully predicts the formation of large scale structure. The general form of the halo is expected to be independent of the total mass enclosed over many orders of magnitude. A useful parametrization that fits the results of the simulations is
\begin{equation}
\rho(r) = \frac{\rho_0}{(r/r_s)^\gamma[1+(r/r_s)^\alpha]^{(\beta-\gamma)/\alpha}},
\label{eq:fitting}
\end{equation}
where among the variables $(\alpha, \beta, \gamma, r_s)$, $\gamma$ is the inner cusp index and $r_s$ the scale radius. In Fig.~1, we plot some commonly used profiles, the Navarro-Frenk-White (NFW)~\cite{Navarro:1995iw}, Moore~\cite{Moore:1999gc}, and Kravtsov~\cite{Kravtsov:1997dp} profiles, with the corresponding parameters listed in Table~I. The normalizations are chosen so that the mass contained within the solar circle ($R_{sc}=8.5 \,\mathrm{kpc}$) provides the appropriate DM contribution to the local rotational curves (see~\cite{Bergstrom:1997fj} for details of a similar analysis). This yields the local DM density $\rho_{NFW}(R_{sc}) = 0.3~\mathrm{GeV}\,\mathrm{cm}^{-3}$ for the NFW profile; for the others, see Table I. 

In general, the profiles agree on larger scales. However, the profiles in the inner regions are more uncertain, due to the spatial resolution limit of the simulations. Physically, baryons typically dominate these regions, and their impact has not yet been fully taken into account in the simulations. The effect of baryons is complicated and can increase the inner cusp strength~\cite{Prada:2004,Gnedin:2004cx} as well as provide a mechanism for momentum transfer to DM particles, which works to flatten cusps~\cite{Mashchenko:2006dm}. With these points in mind, we choose the halo profiles shown in Fig.~1 to illustrate the impact of variations in the halo parametrization on the signal predictions.


\subsection{Annihilations in the Halo}
The calculation of the spectrum of DM annihilation products from the halo is well established (e.g., Ref.~\cite{Bergstrom:1997fj}).  For annihilations, the intensity (number flux per solid angle) at an angle $\psi$ with respect to the GC direction is proportional to the line of sight integration of the DM density squared, or,
\begin{equation}
{\cal J}(\psi) = \frac{1}{R_{sc} \rho_{sc}^2} \int_{0}^{\ell_{max}} \rho^2 ( \sqrt{R_{sc}^2-2\, l\, R_{sc}\cos\psi+l^2} ) \, d\ell,  
\end{equation}
which we choose to normalize at the solar circle ($R_{sc}=8.5 \,\mathrm{kpc}$) with $\rho_{sc} = 0.3$ GeV cm$^{-3}$. Note that the prefactor $1/(R_{sc} \rho_{sc}^2)$ is an arbitrary scaling that is used to make ${\cal J}$ dimensionless regardless of the underlying DM density profile. The upper limit of the integration, 
\begin{equation}
\ell_{max}= \sqrt{(R_{MW}^2-\sin^2 \psi R_{sc}^2)} + R_{sc} \cos \psi \,,
\end{equation}
depends on the adopted size of the halo, $R_{MW}$. However, contributions beyond the scale radius, typically about 20--30 kpc, are negligible. It is also useful to define the average of ${\cal J}$ in a cone with half-angle $\psi$ around the GC that spans a field of view of $ \Delta \Omega = 2 \pi (1 - \cos{\psi}) $:
\begin{equation}
{\cal J}_{\Delta \Omega} = \frac{1}{\Delta \Omega} \int_{\cos \psi}^{1} {\cal J}(\psi^\prime) \, 2 \pi \, d(\cos \psi^\prime)\,.
\end{equation}
Then the average intensity of  DM annihilation products from the field of view can be cast as
\begin{eqnarray}
\frac{d\Phi_{\Delta \Omega}}{dE}= \frac{\langle\sigma_A v\rangle}{2}
{\cal J}_{\Delta \Omega} \frac{R_{sc} \rho_{sc}^2  } {4\pi m_{\chi}^2} \frac{dN}{dE},
\end{eqnarray}
where $dN/dE$ is the spectrum of annihilation products and $m_{\chi}$ is the assumed mass of the DM particle. The factor 1/2 accounts for DM being its own antiparticle and $1/4 \pi$ is for isotropic emission. The thermal average of the annihilation rate is proportional to $\langle\sigma_A v\rangle$, the product of the annihilation cross section and the relative velocity.

In Fig.~2, we show the behavior of ${\cal J}(\psi)$ (thin lines) and ${\cal J}_{\Delta \Omega}$ (thick lines), as functions of $\psi$ and $\Delta \Omega/4\pi$, respectively, for the three representative halo profiles. Note that ${\cal J}$ diverges for very cuspy profiles. This cusp may be an artifact of simulations, which cannot resolve small spatial scales. To avoid such numerical divergences, we put in a flat core for all the profiles in at the innermost $0.1^\circ$, which corresponds to $\simeq$ 0.015~kpc; our results are not sensitive to how the inner profile is regulated.

\begin{figure}[t]
\includegraphics[width=3.25in,clip=true]{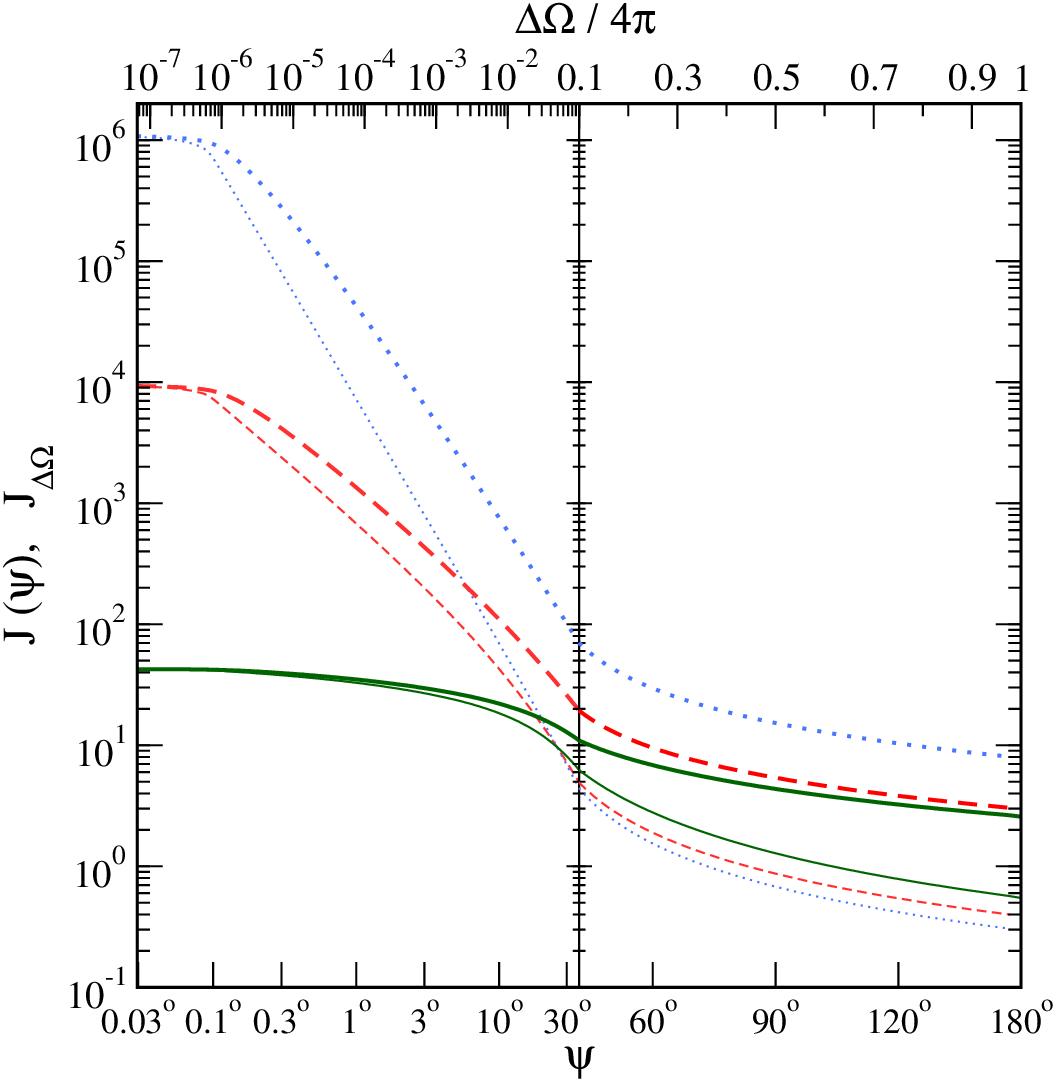}
\caption{Line of sight integration ${\cal J}(\psi)$ as a function of the pointing angle $\psi$ (bottom axis) with respect to the Galactic Center direction for the three different profiles considered (Moore, NFW and Kravtsov in order of dotted, dashed and solid thin lines). Its average, ${\cal J}_{\Delta \Omega}$, inside a cone with half angle $\psi$ around the GC as a function of the visible fraction of the whole sky, $\Delta \Omega/4\pi=(1-\cos{\psi})/2$ (top axis), is also presented (thick lines). Note that the left (right) side of the graph is presented in log (linear) scale in $\Delta \Omega/4\pi$.
\label{fig-2}}
\end{figure}

One can expect very large enhancements for small angles $\psi$ around the GC. Many studies have therefore focused on emission from the few degrees around the GC. However, while this takes advantage of the higher DM densities around the GC, it also strongly depends on the chosen profile. As shown in Fig.~2, the differences in the intensity due to different choices of profile can be orders of magnitude at small angles. We therefore focus on the signal arising within angles $\psi \simeq 30^\circ$ or larger around the GC. This is motivated for three reasons. First, the enhancement $ {\cal J}_{\Delta \Omega}$ is much less sensitive to the chosen profile for larger $\psi$. Second, the viewing angle $\psi$ to consider depends on the particle considered, and for neutrinos, the angular resolution is limited by the kinematics of the detection reactions to be $\delta \theta(E) \simeq 30^\circ \times{\sqrt{[GeV]/E}}$. Thus, it is impractical to consider the arrival direction for neutrinos less energetic than several GeV. Third, choosing a large field of view ensures that there are adequate statistics for the detection of the signal and background.

We now define three signals corresponding to different components of the halo:
\begin{itemize}
\item \emph{Halo Angular:} We consider a 30$^\circ$ half-angle cone for neutrinos more energetic than 10 GeV and a larger half-angle of $\simeq3 \times \delta \theta(E)$ for lower energies.  A conservative value for this \emph{Halo Angular} component inside $\psi\simeq 30^\circ$ is ${\cal J}_{Ang}\simeq 25$ (below 10 GeV, when larger angular regions are used, this would be smaller). 
\item \emph{Halo Average:} When a large field of view $ \Delta \Omega$ is considered, the average ${\cal J}_{\Delta \Omega}$ is much less sensitive to the chosen profile, as shown by the differences in
${\cal J}_{\Delta \Omega}$ in Fig.~\ref{fig-2}.  We define the \emph{Halo Average}, ${\cal J}_{Ave} \simeq 5 $, as a canonical value for the whole sky ($\psi = 180^\circ$).  Note that the differences due to different choices of profiles from this canonical value are small, less than a factor of 2. 
\item \emph{Halo Isotropic:} Another key point is that ${\cal J}(\psi)$ never vanishes, even in the anti-GC direction ($\psi=180^\circ$), due to a minimal isotropic component of the halo signal that arises within our immediate neighborhood. This is present for all profiles and cannot be differentiated from the truly cosmic signal, except by its spectrum.  We define this minimal component as \emph{Halo Isotropic}, ${\cal J}_{Iso} \simeq 0.5  $, which yields the most robust constraints. 
\end{itemize}
The values for these three components are summarized in Table~I.  Finally, halos may have substructure which can contribute an additional boost to DM annihilation signals, e.g., Refs.~\cite{Hooper:2007be,Silk:1992bh,Bi:2005im}. To be conservative, we do not attempt to model any substructure, but note that it would only enhance all of the annihilation signals and hence improve the constraints that we present below. 


\subsection{Cosmic Annihilations}
The calculation of the diffuse cosmic DM annihilation signal has been thoroughly discussed in the literature (e.g., Ref.~\cite{Ullio:2002pj}). The intensity of the cosmic diffuse signal is 
\begin{equation}
\frac{d\Phi}{dE} = 
\frac{\langle \sigma_A v\rangle}{2}
\frac{\Omega_{\chi}^2\rho_{\rm c}^2}{4 \pi m_\chi^2}
\frac{c}{H_0}
\int \frac{dN(E')}{dE'}
\frac{(1+z)^3 f(z)}{h(z)} \, dz\,,
\label{cosmicdiffuse}
\end{equation}
where $\rho_{\rm c}= 5.3 \times 10^{-6}~\mathrm{GeV}\,\mathrm{cm}^{-3} $ is the critical density, $H_0=70 \, \text{km} \, \text{s}^{-1} \, \text{Mpc}^{-1}$ is the Hubble constant, and $h(z)= [(1+z)^3 \Omega_{\chi} + \Omega_\Lambda]^{1/2}$ with $\Omega_{\chi}=0.3$ (ignoring the small baryonic contribution) and $\Omega_\Lambda=0.7$. The original spectrum of annihilation products is redshifted through $E^\prime = (1+z) E$, yielding $dN(E')/dE'$. For gamma rays above a few TeV, pair production with the diffuse extragalactic  infrared and optical photon backgrounds is an important loss mechanism. With neutrinos, such attenuation considerations are unnecessary. (Attenuation in Earth would be important for neutrinos above about 100 TeV, e.g., Ref.~\cite{attenuation}, mostly beyond our range of interest.)

The most relevant factor is the cosmic source intensity multiplier, $f(z)$, that reflects the formation history of halos. This is a decreasing function of increasing redshift, and although the redshift dependence is fairly universal, the normalization varies quite considerably, by more than an order of magnitude for different profiles~\cite{Ando:2005hr}. The impact of its evolution on the predicted cosmic signal is milder once the overall normalization is determined. Thus we parametrize it with a simple fitting form
\begin{equation}
f(z)=f_0 \times 10^{ 0.9 (\exp[-0.9 z]-1) - 0.16z } \,,
\label{evolution}
\end{equation}
where $f_0$ sets the overall normalization. This normalization is determined by the underlying halo profile, concentration parameter, and inclusion/exclusion of either smaller halos or subhalos, and its dependence on the halo profile alone can be seen in Table~I when the other parameters are fixed. The uncertainties of the halo signal (including the inner cusp profile, whether a spike or core has formed, and substructure), are important for the cosmic signal predictions as well. The cosmological clustering of halos, halo mass function and lower mass cutoff are additional uncertainties associated with the cosmic signal. Thus, the halo signal is overall less uncertain, only factors of a few, compared to the cosmic signal, which varies by more than an order of magnitude for different profiles; compare the entries in Table~I.

\begin{table}[t]
\caption{The parameters of Eq.~(\ref{eq:fitting}) for the three profiles considered. Scale radius, $r_s$, is in [kpc] and the normalization at solar circle, $\rho(R_{sc})$, is in units of [GeV cm$^{-3}$]. Estimates of the various enhancement factors, ${\cal J}$, for the halo (as defined in the text) and the normalization of the cosmic intensity multiplier, $f_0$, in units of [10$^5$] are also summarized. The last row lists canonical values that we use to derive our constraints on the DM total annihilation cross section.}
\begin{ruledtabular}
\begin{tabular}[t]{lccccccccc}
{\extracolsep{\fill}}&$\alpha$ & $\beta$ & $\gamma$ & $r_s$ & $\rho(R_{sc})$ &${\cal J}_{Ang}$ & ${\cal J}_{Ave}$ & ${\cal J}_{Iso}$ & $f_0$ \\
\hline
Moore & 1.5& 3 & 1.5& 28 & 0.27 & 102 & 8 & 0.3 & 5 \\
NFW & 1& 3 & 1 & 20 & 0.3 & 26 & 3 & 0.4 & 0.5 \\
Kravtsov & 2& 3& 0.4 & 10 & 0.37 & 13 & 2.6 & 0.55 & 0.2 \\
\hline
{\bf Canonical} & & & & & & {\bf25} & {\bf5} & {\bf0.5} & {\bf1}\\
\end{tabular}
\end{ruledtabular}
\end{table}

The ratio of the DM annihilation halo signal to the cosmic signal can be approximated as
\begin{eqnarray}
\frac{\Phi^{H}_{\Delta \Omega}}{\Phi^{C}}\sim
\frac{{\cal J}_{\Delta \Omega} R_{sc} \rho_{sc}^2}{
c H_0^{-1} \Omega_{\chi}^2\rho_{\rm c}^2 f_0} 
\sim 10^5 \, \frac{{\cal J}_{\Delta \Omega}}{f_0},
\end{eqnarray}
which is independent of $\langle \sigma_A v\rangle$ and $m_\chi$. Now, using the \emph{Halo Isotropic} signal (${\cal J}_{Iso} \simeq 0.5 $) for illustration, this ratio will be close to unity for $f_0 \simeq 5 \times 10^4$ (coincidentally, equal to the NFW value as reported in Table~I). Therefore, even the \emph{Halo Isotropic} enhancement is significant. Its contribution to the apparently diffuse signal observed at Earth is comparable to that of the truly cosmic signal for profiles as flat as NFW, and it exceeds the cosmic signal for profiles flatter than the NFW. This was also pointed out in Ref.~\cite{Oda:2005nv}, specifically for the NFW profile with substructures.

The flatter profiles are well-motivated both theoretically and observationally, and hence the \emph{Halo Isotropic} can be an important component. While simulations tend to show the presence of a mild cusp, there are also observations indicating otherwise, motivating a flatter, cored profile~\cite{Gentile:2004tb}. It is not known whether a core is a general feature of the halo structure. In contrast to flatter profiles, more cuspy profiles like the Moore profile yield much larger fluxes for both the \emph{Halo Average} and the cosmic signal, and then the \emph{Halo Isotropic} will be relatively insignificant. In our calculations below, we use a fairly large value of $f_0\simeq 10^5$ to be conservative in comparing constraints that can be obtained from the halo or cosmic annihilations. It is unlikely that $f_0$ can be any larger, unless contributions from the halo substructure is significant, which will boost the halo signal as well, not significantly modifying their relative strength. Constraints from the halo will be even more important for smaller $f_0$ values.

While we use neutrinos for constraining the total annihilation cross section, the gamma-ray flux has been used in many past studies. In this regard, a recent study by Ando~\cite{Ando:2005hr}, which compared the DM annihilation gamma-ray signal from the GC to the cosmic signal is of interest. It concluded that under the assumptions that annihilations account for the gamma ray flux from the GC and that halo profiles are universal, then annihilations cannot be the dominant component of the isotropic extragalactic gamma ray background. That study dealt with halos at least as steep as NFW; the \emph{Halo Isotropic}, on the other hand, is most interesting for flatter profiles, as we have shown. The \emph{Halo Isotropic} would increase the `apparently' (the cosmic signal plus the isotropic halo component) extragalactic gamma ray background, without violating the constraints placed using the GC data. 


\section{Neutrino Constraints}
To obtain the most robust limits on the total annihilation cross section for dark matter, we assume that annihilations produce only neutrinos (the least detectable known particles) and compare the resultant neutrino signal to the atmospheric neutrino measurements.  We adopted a simple model in which dark matter particles annihilate into  pairs of neutrinos, $\chi +\chi \rightarrow \bar{\nu}+ \nu$, and the spectrum of the neutrinos per flavor is a monochromatic line with ${dN_\nu}/{dE} = \frac{2}{3} \delta(E-m_{\chi}) $, where $m_{\chi}$ is the mass of the dark matter particle. The prefactor 2/3 arises under the assumption that 2 neutrinos are produced per annihilated DM pair and that all neutrino flavors are equally populated (either in production or through neutrino mixing). This delta function is regularized by the redshift integration for the cosmic signal. For the sake of a simple graphical representation in Fig.~3, we used a narrow Gaussian (width $\epsilon = m_{\chi}/20 $) to represent the line signal.

Strictly speaking, it is not possible to have {\it only} neutrinos in the final state, due to electroweak radiative corrections that lead to $Z$ and $W$ bremsstrahlung~\cite{Berezinsky:2002hq}.  While these corrections do not significantly affect our neutrino flux calculations, the vector-boson decay chains do lead to gamma-ray fluxes.  In turn, these provide a new way to model-independently limit the DM total annihilation cross section with results comparable to those obtained with the cosmic neutrino fluxes~\cite{Kachelriess-Serpico}.

\begin{figure}[t]
\includegraphics[width=3.25in,clip=true]{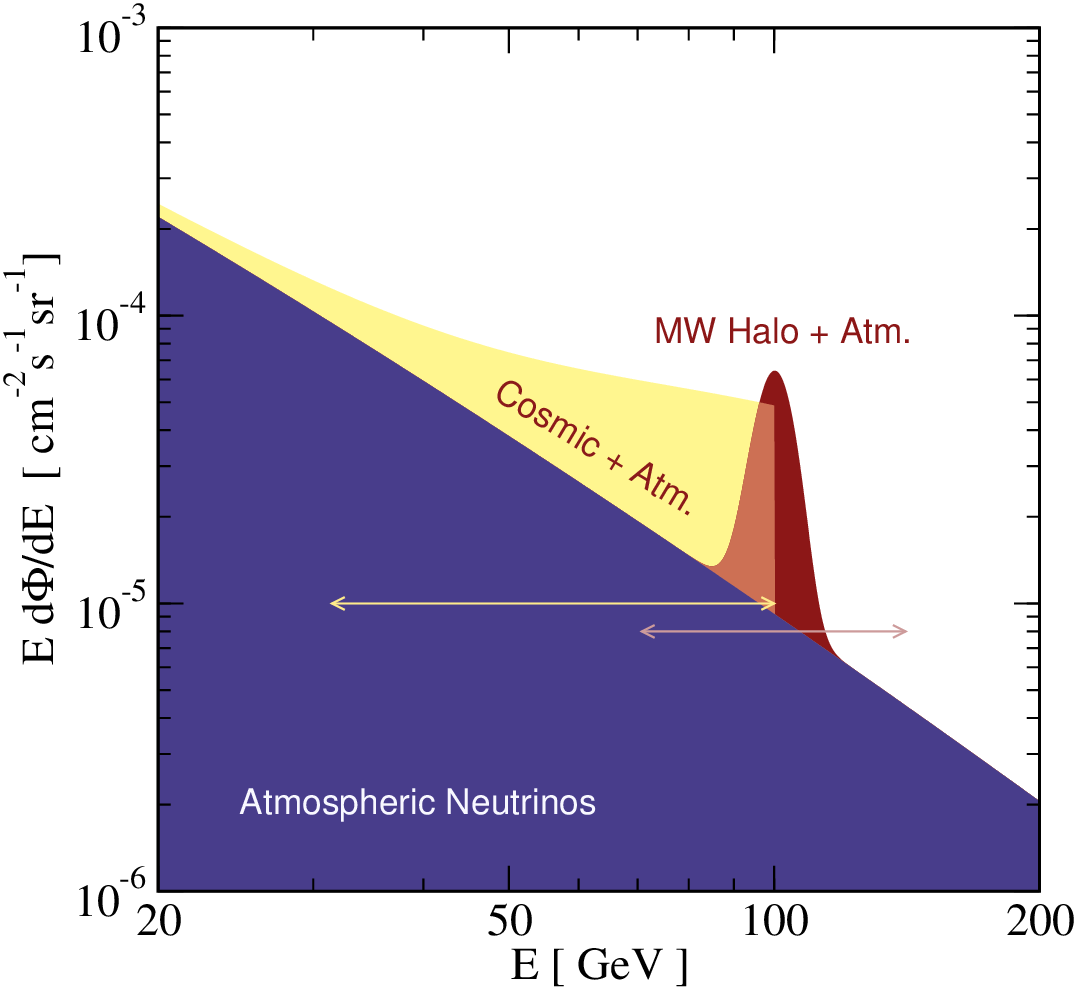}
\caption{The atmospheric neutrino spectrum ($\nu_{\mu} + \bar{\nu}_{\mu}$) and the expected increase in the total received intensity from the annihilations of a $100$~GeV DM particle. While the halo signal is sharply peaked in a narrow band of energy, the cosmic signal spreads over a wider energy range.  To define upper limits on the cross section, in each case we use a (different, see Fig.~4) cross section such that the annihilation signal would double the total received intensity in the displayed energy ranges. 
\label{fig-3}}
\end{figure}

In Fig.~\ref{fig-3}, we illustrate how we set our conservative bound on the total annihilation cross section using neutrinos. The atmospheric neutrino spectrum (specifically, $\nu_\mu + \bar{\nu}_\mu$) is derived from measurements by the Fr\'ejus, Super-Kamiokande and AMANDA detectors, for which data are available up to $10^{5}$~GeV~\cite{atmospheric}. The primary source is the decay of pions and kaons produced by cosmic ray collisions with the atmosphere. While the intensities depend on zenith angle, we adopted an average value which will suffice for our purposes (see Ref.~\cite{Beacom:2006tt} for details).  In the full energy range considered, and in fact to higher energies, all of the data are reported as being consistent with the theoretically expected flux and energy spectrum shape for atmospheric neutrinos.  There is no evidence of any other signals, e.g., from astrophysical sources.

The DM annihilation signals are calculated using a fiducial DM mass of $m_{\chi} = 100$~GeV ($f_0\simeq 10^5$ for the cosmic signal and ${\cal J}_{Iso} \simeq 0.5 $ for the \emph{Halo Isotropic} signal) and are superimposed on the atmospheric neutrino spectrum (per flavor, using $\nu_\mu + \bar{\nu}_\mu$). Since the halo signal is sharply peaked, we chose an energy bin width of $\Delta \log_{10}{E} = 0.3$  around $E=m_{\chi}$ to compare it to the atmospheric neutrinos, require it to double the total received intensity in this energy bin, and adjust the cross section accordingly. The additional smearing due to redshifting necessitates a larger bin width of $\Delta \log_{10}{E} = 0.5$  below $E=m_{\chi}$ for the cosmic signal. These choices are within the energy resolution limits of the neutrino detectors~\cite{Beacom:2006tt,atmospheric}.

\begin{figure}[t]
\includegraphics[width=3.25in,clip=true]{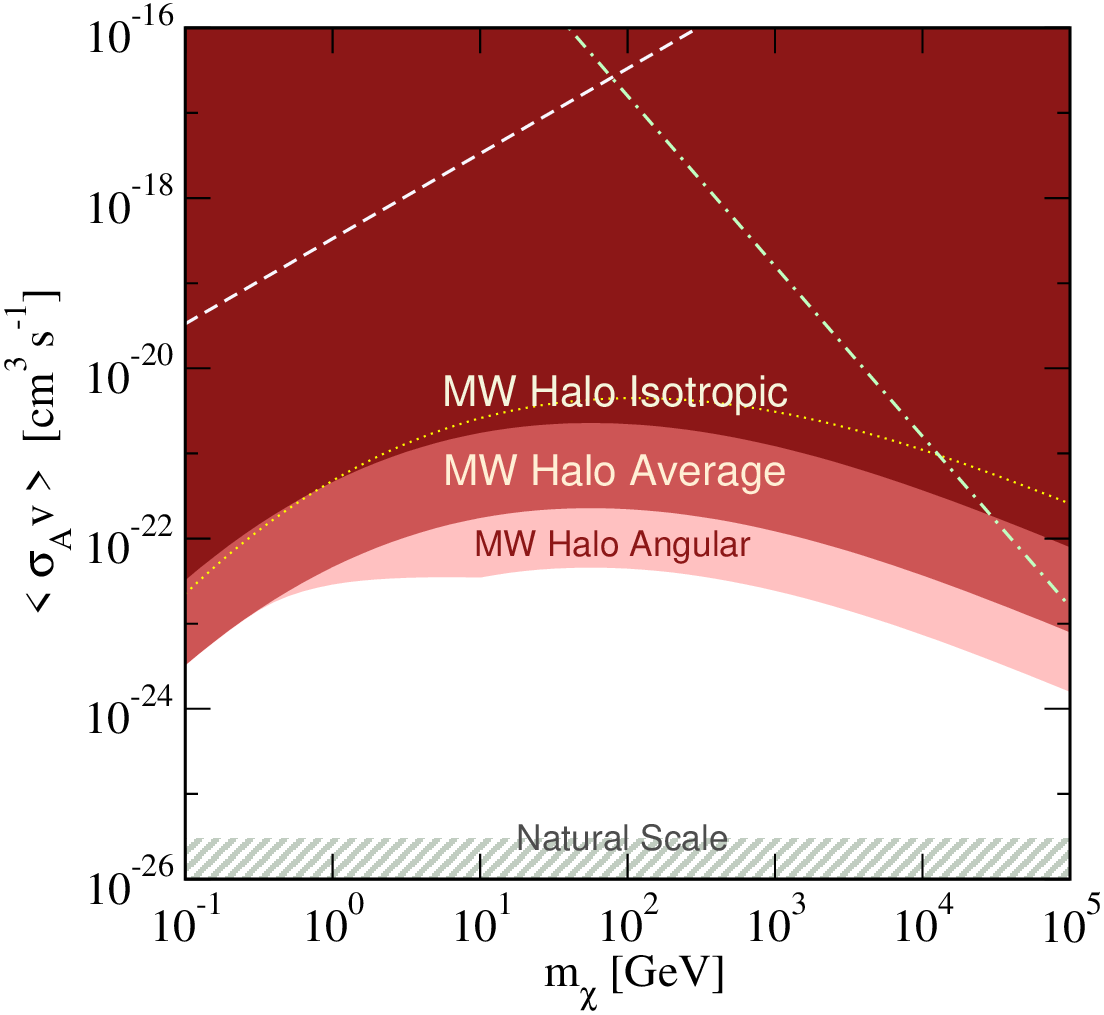}
\caption{Constraints on the DM total self-annihilation cross section from various components of the Milky Way halo (shaded regions excluded). The constraints from the cosmic signal (dotted line) can be compared to that of the \emph{Halo Isotropic} (dark shaded) component. The successive improvements of the \emph{Halo Average} (medium shaded) and  \emph{Halo Angular} (light shaded) components can improve on the cosmic constraint by 1--2 orders of magnitude. The unitarity bound (dot-dashed line) and the cross sections for which annihilations flatten the cusps of DM halos (dashed-line) are also shown; see the text.
\label{fig-4}}
\end{figure}

The cross section limits obtained from the above analysis are presented in Fig.~\ref{fig-4} as a function of the DM mass.  We show results derived using three different components of the halo: the \emph{Halo Angular}, the \emph{Halo Average} and the \emph{Halo Isotropic} (shaded regions in order from lighter to darker, with each also excluding all larger cross section values), and also the cosmic diffuse signal (dotted line). The enhancement factors used in each case are listed in Table~I. Any larger cross section would produce too much intensity and cannot be accommodated by the data. We also show the unitarity bound (dot-dashed line)~\cite{Griest:1989wd,Hui:2001wy, Beacom:2006tt}, and the range of cross sections for which DM annihilation may flatten cusps (dashed line)~\cite{Kaplinghat:2000vt, Beacom:2006tt}; cross sections above the lines are excluded.  The natural scale of the DM annihilation cross section (hatched region at the bottom) is shown for a thermal relic.  However, larger cross sections (not exceeding the present limits) are possible, e.g., Ref.~\cite{Kaplinghat:2000vt}, if dark matter is produced non-thermally or acquires mass only in the late universe.  (For such nonstandard dark matter scenarios, it is also possible for the DM to have a large scattering cross section with baryons, though this is now strongly constrained, e.g., Refs.~\cite{ScatteringLimits}.)

The \emph{Halo Isotropic} component, which appears in all directions, is not distinguishable (apart from its spectrum) from the cosmic signal, and provides the most robust constraint. We also consider the \emph{Halo Average} and  \emph{Halo Angular} components that are relatively more directional, yet they can improve on the cosmic constraint by 1--2 orders of magnitude due to their larger signal intensities. If the inner density profile were assumed to be known, a greater enhancement in the neutrino signal could be possible by restricting to a smaller angular region, as in Ref.~\cite{Bertone:2004ag}.   The actual neutrino experiments  have different exposures to different parts of the halo, which should be taken into account in a more thorough analysis. Thus our actual constraint that is indicated by the data is most faithfully represented by the \emph{Halo Average}. 

Several advantages of the halo signal over the cosmic signal can be summarized as follows:
\begin{itemize}
\item While both suffer from uncertainties such as the concentration parameter and the shape of the halo, the halo signal on large scales is overall more certain, as noted, and shown in Table~I. 
\item The isotropic component of the halo signal is especially important for flatter profiles, for which it dominates over any truly cosmic signal. For cuspy profiles, the \emph{Halo Angular} would be even more constraining than displayed in our Fig.~4.
\item The cosmic signal is broadened in energy by redshifting, making it harder to identify over the smoothly varying atmospheric neutrino spectrum. 
\item Gamma rays from cosmic DM annihilations are attenuated at high energies, thus the statement that ``anything other than neutrinos will be more detectable'' may not be always fully applicable for the cosmic signal (the halo signal will still be present). 
\end{itemize}
The neutrino limits derived from the halo signal in this study not only improve on the previous limit derived from the cosmic signal, but also extend their applicability.


\section{Conclusions}
Can neutrinos truly be the only final state? Regardless of the answer, they can yield the most conservative limit on the DM  total self-annihilation cross section~\cite{Beacom:2006tt}.  Neutrinos are the least detectable particles of the Standard Model and all other final states produce more visible signatures. Thus a limit on their flux (assuming DM only annihilates into neutrinos) conservatively bounds the total annihilation  cross section of dark matter. We focus on the importance of the DM annihilation signal arising from the halo, while BBM used the truly cosmic DM annihilation signal (for decaying DM scenarios see Ref.~\cite{Yuksel:2007dr} and references therein). We compare the expected spectra of the neutrinos from annihilations to the well-measured atmospheric neutrino spectrum and arrive at our constraints on the DM self annihilation cross-section. We show that, by using the halo signal, the previous bounds from the cosmic signal are improved on by 1--2 orders of magnitude.

This illustrates the point that our intermediate results are more generally applicable than just for neutrinos. The halo signals have only minor model dependence compared to the GC and cosmic signals, and also they can be significantly larger than the cosmic signals, particularly for flatter profiles.  These results will also be useful for studies with gamma rays.

An additional enhancement due to possible sub-structure of the dark matter or dedicated analyses of the data~\cite{Beacom:2006tt} may bring this improvement up to 2--4 orders of magnitude. This may turn neutrinos into surprisingly effective tools for exploring the full parameter space of the total annihilation cross section and mass of the DM, perhaps even testing the cross-section scale of thermal relics.  We emphasize that in the most interesting mass range (for lower masses, see Ref.~\cite{PalomaresRuiz:2007eu}), the cosmic and halo neutrino constraints are already significantly improving on the unitarity bound~\cite{Griest:1989wd, Hui:2001wy}.  Our bounds are especially useful at energies $>100$ GeV in which there are no gamma-ray data on large angular scales, whereas the neutrino data go to $> 100$ TeV.

In the short term, we strongly encourage dedicated analyses of these halo signals by the Super-Kamiokande and AMANDA/IceCube collaborations.  Our requirement that the DM neutrino signal be 100\% as large as the atmospheric neutrino background in specified and relatively large energy and angular ranges is too conservative.  The atmospheric neutrino background has been measured with high statistics over six orders of magnitude in energy, and generally, only much smaller perturbations would be allowed.  Together with the other strategies for improving sensitivity noted in Ref.~\cite{Beacom:2006tt}, we expect that dedicated analyses will lead to much more stringent limits than estimated here.

In the coming age of near-term and proposed super-sized detectors, like GLAST, IceCube, LHC and Hyper-Kamiokande, together with low background underground facilities with unprecedented precision, the days of DM as an elusive and hidden substance may finally be coming to a conclusion. Neutrinos may play a key role in unveiling the true identity of the dark matter. We have demonstrated the excellent prospects of neutrino detectors as the largest but most conservative tools to constrain the DM total annihilation cross section and mass.


\section*{Acknowledgments}
We thank N.~Bell, M.~Kamionkowski, M.~Kistler, G.~Mack, E.~Rozo, W.~Wada, and C.~Watson for helpful comments. HY and JFB are supported by NSF CAREER Grant No.~PHY-0547102 to JFB, and by CCAPP at the Ohio State University.  SA is supported by the Sherman Fairchild Foundation.


\end{document}